\documentclass[%
 reprint,
 amsmath,amssymb,
 aps,
]{revtex4-2}

\usepackage{graphicx}
\usepackage{dcolumn}
\usepackage{bm}
\usepackage{hyperref}
\usepackage{multirow}
\usepackage{amsfonts}
\usepackage{xcolor}

\begin{document}

\preprint{APS/123-QED}

\title{Topological electron and phonon flat bands in novel kagome \\superconductor $X$Pd$_{5}$ ($X$=Ca, Sr, Ba)}

\author{Jiefeng Ye$^1$}
\author{Zhigao Huang$^1$}

\author{Xianxin Wu$^2$}
 \email{xxwu@itp.ac.cn}

\author{Jian-Min Zhang$^1$}%
 \email{jmzhang@fjnu.edu.cn}

\affiliation{%
 $^1$Fujian Provincial Key Laboratory of Quantum Manipulation and New Energy Materials, College of Physics and Energy,\\Fujian Normal University, Fuzhou 350117, China
 }%
\affiliation{%
 $^2$Institute of Theoretical Physics, Chinese Academy of Sciences, Beijing 100190, China
 }%

\date{\today}

\begin{abstract}
  Fermionic and bosonic localized states induced by geometric frustration in the kagome lattice provide a distinctive research platform for investigating emergent exotic quantum phenomena in strongly correlated systems. Here, we report the discovery of coexisting electronic and phononic flat bands induced by geometric frustration in a novel kagome superconductor $X$Pd$_{5}$ ($X$=Ca, Sr, Ba). The electronic flat band is located around the Fermi level and possesses a nontrivial topological invariant with $\mathbb{Z}_2 = 1$. Additionaly, we identify multiple van Hove singularities (vHS) arise from the kagome Pd $d$ orbitals with distinct dispersion and sublattice features, including conventional, higher-order vHS and p-type, m-type vHS. Specifically, our investigation of the vibrational modes of the phononic flat band reveals that its formation originates from destructive interference between adjacent kagome lattice sites with antiphase vibrational modes. A spring-mass model of phonons is established to probe the physical mechanism of the phononic flat bands. Furthermore, the calculations of electron-phonon coupling in the $X$Pd$_{5}$ reveal superconducting ground states with critical temperatures ($T_c$) of 4.25 K, 2.75 K, and 3.35 K for CaPd$_{5}$, SrPd$_{5}$, and BaPd$_{5}$, respectively. This work provides a promising platform to explore the Fermion–boson many-body interplay and superconducting states, while simultaneously establishing a novel analytical framework to elucidate the origin of phononic flat bands in quantum materials.
\end{abstract}

\maketitle


\section{INTRODUCTION}

\begin{figure}
  \includegraphics[width=\columnwidth]{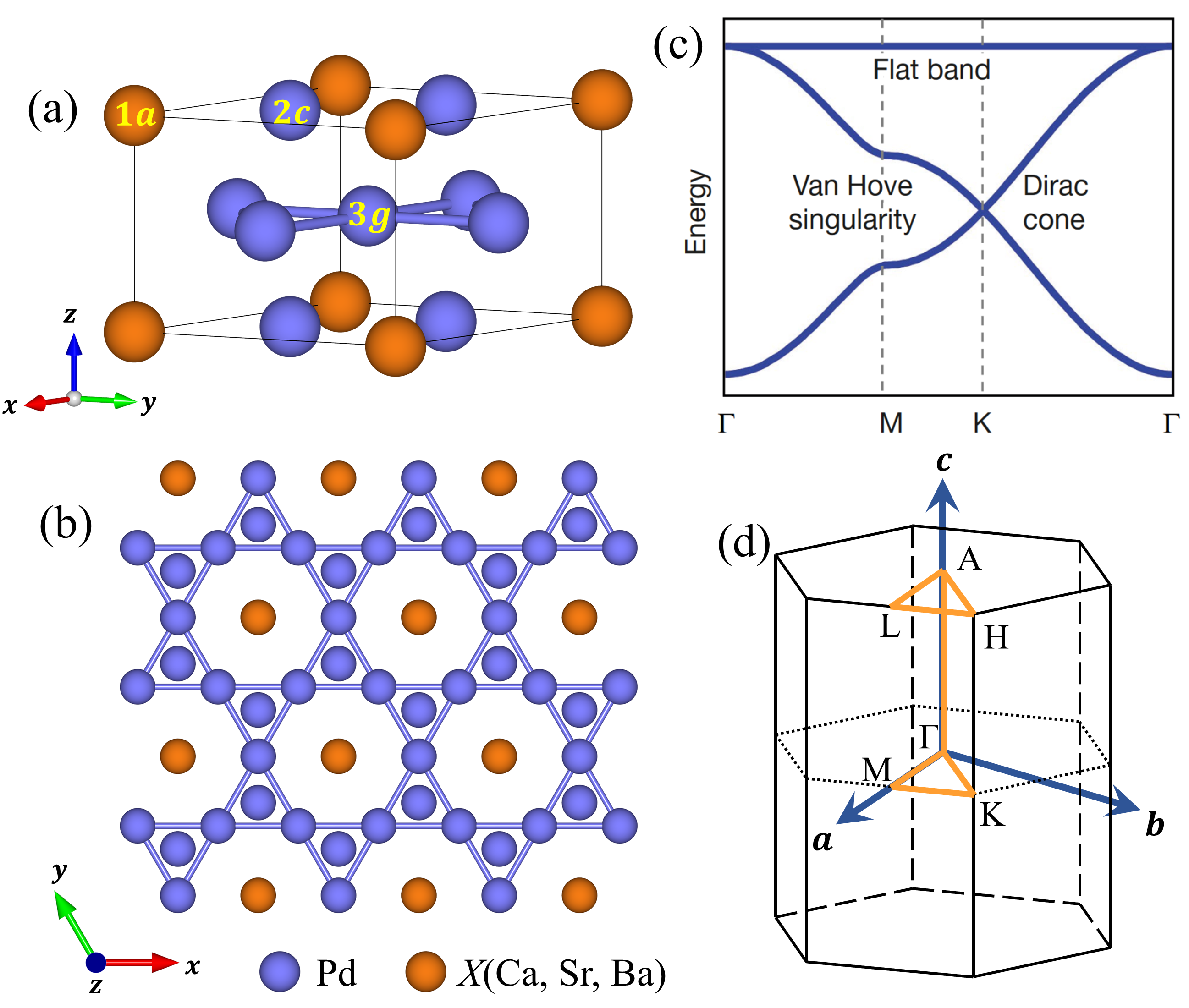}
  \caption{\label{fig1} Crystal structure of kagome lattice $X$Pd$_5$ in (a) side view and (b) top view. (c) Tight-binding model of a kagome lattice considering the nearest-neighbour electron hopping term\cite{yin2022topological}. (d) Brillouin zone diagram of $X$Pd$_5$ with labeled high-symmetry points. }\end{figure}

It is well known that the unique quasi-two-dimensional frustrated structure of the kagome lattice induces destructive interference in subdimensional eigen-wavefunctions, leading to the quenching of quantum kinetic energy, localized states, and the emergence of dispersionless flat bands (FB). Besides, the hexagonal sublattice produces Dirac point (DP) at the Brillouin corners and two van Hove singularities (vHS) at boundaries, as illustrated in Fig. \ref{fig1}(c). These distinctive features of the kagome lattice represent unique quantum behaviors that will induce a series of novel physical phenomena, including fractional quantum Hall effect\cite{tang2011high,neupert2011fractional,lu2024fractional,kang2024evidence}, Fermi liquid\cite{tazai2023rigorous,ye2024hopping}, unusual charge order\cite{nie2022charge,zheng2022emergent,teng2023magnetism,hu2024phonon} and unconventional superconductivity\cite{cao2018unconventional,wu2021nature,ortiz2020cs,liu2024superconductivity,li2025mpd}.

One widely studied superconducting ground state in the kagome lattice is characterized by the presence of a flat band at the Fermi level ($E_{\mathrm{F}}$), which has been experimentally proven to exhibit unconventional superconductivity\cite{cao2018unconventional,liu2024superconductivity,wu2025flat}. The interplay between nontrivial topological electronic band structures and the emergence of intrinsic superconducting states renders this field a particularly compelling platform for realizing exotic ground states and quasiparticles. Furthermore, the kagome lattice can also host topological phononic flat bands, where nontrivial many-body interactions may emerge between lattice phonons and electrons\cite{yin2020fermion}. Experimental and theoretical evidence indicates that the formation of charge density waves (CDW) correlates with the presence of phononic flat bands\cite{hu2025flat,cao2023competing,korshunov2023softening}. However, despite extensive theoretical model studies on topological phonon flat bands\cite{xia2024negative}, research on their realization in real materials remains limited, and the mechanism by which phonons form flat bands under the influence of geometric frustration remains unclear. Therefore, identifying a material that simultaneously hosts topological electronic flat bands and phonon flat bands is crucial for uncovering the underlying mechanisms of these novel physical phenomena. Unfortunately, due to the influence of diverse symmetries and complex electronic degrees of freedom in real crystals, ideal characteristics are extremely scarce in kagome materials.

In this study, we identify $X$Pd$_{5}$ ($X$=Ca, Sr, Ba), a novel class of stable kagome metals that simultaneously exhibits topological electronic flat bands, topological phononic flat bands, and the superconducting ground state. By analyzing the parity of the electronic wavefunctions, we determine that the system is a topological metal with a strong topological index of $\mathbb{Z} _2=1$. Additionally, threefold vHS exist near the Fermi level, originating from the kagome geometry of the Pd $d$ orbital and the parity degrees of freedom, which include two p-type conventional vHS and a m-type higher-order vHS. Typically, phonon flat bands arise from collective lattice displacements within the plane of hexagonal rings on the kagome lattice\cite{yin2020fermion, yin2022topological}. However, in the $X$Pd$_5$ system, we reveal the formation mechanism of the phonon flat bands which are induced by geometric frustration. Specifically, these flat bands originate from out-of-plane vibrations at neighboring sites of the hexagonal ring, where opposite vibrational phases lead to destructive interference, resulting in the emergence of phonon flat bands. Furthermore, we have confirmed the existence of a superconducting state within this topological system via electron-phonon coupling, where the superconducting critical temperatures ($T_c$) are determined to be 4.25 K, 2.75 K, and 3.35 K for CaPd$_{5}$, SrPd$_{5}$, and BaPd$_{5}$, respectively. Our work presents $X$Pd$_{5}$ as a novel topological metal featuring a superconducting ground state, serving as an ideal platform for investigating strong correlations and other emergent physical phenomena.

\section{THEORY AND COMPUTATIONAL DETAILS}

In this work, density functional theory (DFT) calculations of $X$Pd$_{5}$ were performed using the Vienna Ab initio Simulation Package (VASP)\cite{kresse_ab_1993,kresse_efficient_1996} with projector augmented-wave (PAW) potentials\cite{blochl_projector_1994}. The Perdew-BurkeErnzerhof (PBE) function within the generalized gradient approximation (GGA)\cite{perdew_generalized_1996} was adopted to treat the exchange correlation potentials. The electronic calculations used a plane-wave energy cutoff of 300 eV. The $X$Pd$_{5}$ structures were fully relaxed until the residual force per atom was less than 0.01 eV/{\AA}, and the criterion for energy convergence was $10^{-5}$ eV.  The Brillouin zone was sampled with a $\Gamma $-centered grid with a size of $9 \times  9 \times  8$.  The phonon spectrum and atomic displacements were calculated with a $2 \times  2 \times  2$ supercell using density functional perturbation theory (DFPT)\cite{baroni2001phonons} as implemented in the Phonopy code\cite{phonopy-phono3py-JPCM, phonopy-phono3py-JPSJ}. The irreducible corepresentations of states are analyzed by using the IRVSP code\cite{gao2021irvsp}. The edge spectrum is studied by using the WannierTools Package\cite{WU2017}, with the ab-initio tight-binding Hamiltonian constructed by the WANNIER90 package\cite{Mostofi2014}.

As for the superconducting properties, the Quantum Espresso (QE) package\cite{giannozzi2009quantum} was adopted to perform the DFT calculations, in which the PAW pseudopotentials under the PBE scenario\cite{blochl_projector_1994} were generated by PSLIBRARY\cite{DALCORSO2014337}. The cutoff energy of wave functions and charge 
density were set as 60 and 550 Ry, respectively. Phonon dispersion curves were calculated based on DFPT\cite{baroni2001phonons}, in which a denser $12\times 12\times 12$ $k$-point grid and a $6\times 6\times 6$ $q$-point mesh were applied for the electron-phonon coupling (EPC) calculations.

\section{RESULTS AND DISCUSSION}
\subsection{Structures and electronic topological properties}

\begin{table}[b]
  \caption{\label{tab1}%
  Lattice parameters and formation energy($E_f$) of $X$Pd$_{5}$ ($X$=Ca, Sr, Ba).
  }
  \begin{ruledtabular}
  \begin{tabular}{cccc}
  System & $a$ ({\AA}) & $c$ ({\AA}) & $E_f$ (eV) \\ \hline
  CaPd$_5$     & 5.291 & 4.441 & -2.390  \\
  SrPd$_5$     & 5.398 & 4.412 & -2.502  \\
  BaPd$_5$     & 5.555 & 4.334 & -2.375 
  \end{tabular}
  \end{ruledtabular}
\end{table}

The novel paramagnetic kagome lattice $X$Pd$_{5}$ crystallize in the hexagonal crystal system within the space group $P6/mmm$ (No. 191),  exhibiting both time-reversal symmetry and spatial-inversion symmetry. The unit cell consists of two layers stacked along the $c$-axis, with a central kagome layer formed by Pd1 atoms occupying the Wyckoff position $3g$, and adjacent layers composed of Pd2 atoms at the $2c$ site and Ca atoms at the $1a$ site, as shown in Fig. \ref{fig1}(a). Table \ref{tab1} presents the lattice constants and formation energies of the CaPd$_{5}$, SrPd$_{5}$, and BaPd$_{5}$ systems after full relaxation. The formation energy ($E_f$) is calculated using the following formula:
\begin{eqnarray}
  E_{f}=E_{\mathrm{tot}}-\mu_{\mathrm{Ca}}-5 \mu_{\mathrm{Pd}},
\end{eqnarray}
where $E_{\mathrm{tot}}$ represents the Helmholtz free energy of $X$Pd$_{5}$, and $\mu_{\mathrm{Ca}}$ and $\mu_{\mathrm{Pd}}$ denote the chemical potentials of Ca and Pd, respectively. According to the results in Table \ref{tab1}, the $E_{f}$  of the SrPd$_5$ is lower than that of the CaPd$_5$, while the BaPd$_5$ exhibits a formation energy very close to that of the CaPd$_5$, with all values being negative. Previous experimental studies have successfully synthesized CaPd$_{5}$ samples\cite{kilduff2016chemical}, suggesting that the $X$Pd$_{5}$ systems are thermodynamically stable and could potentially be synthesized experimentally by designing appropriate conditions and methods.

\begin{figure*}
  \includegraphics[width=0.9\textwidth]{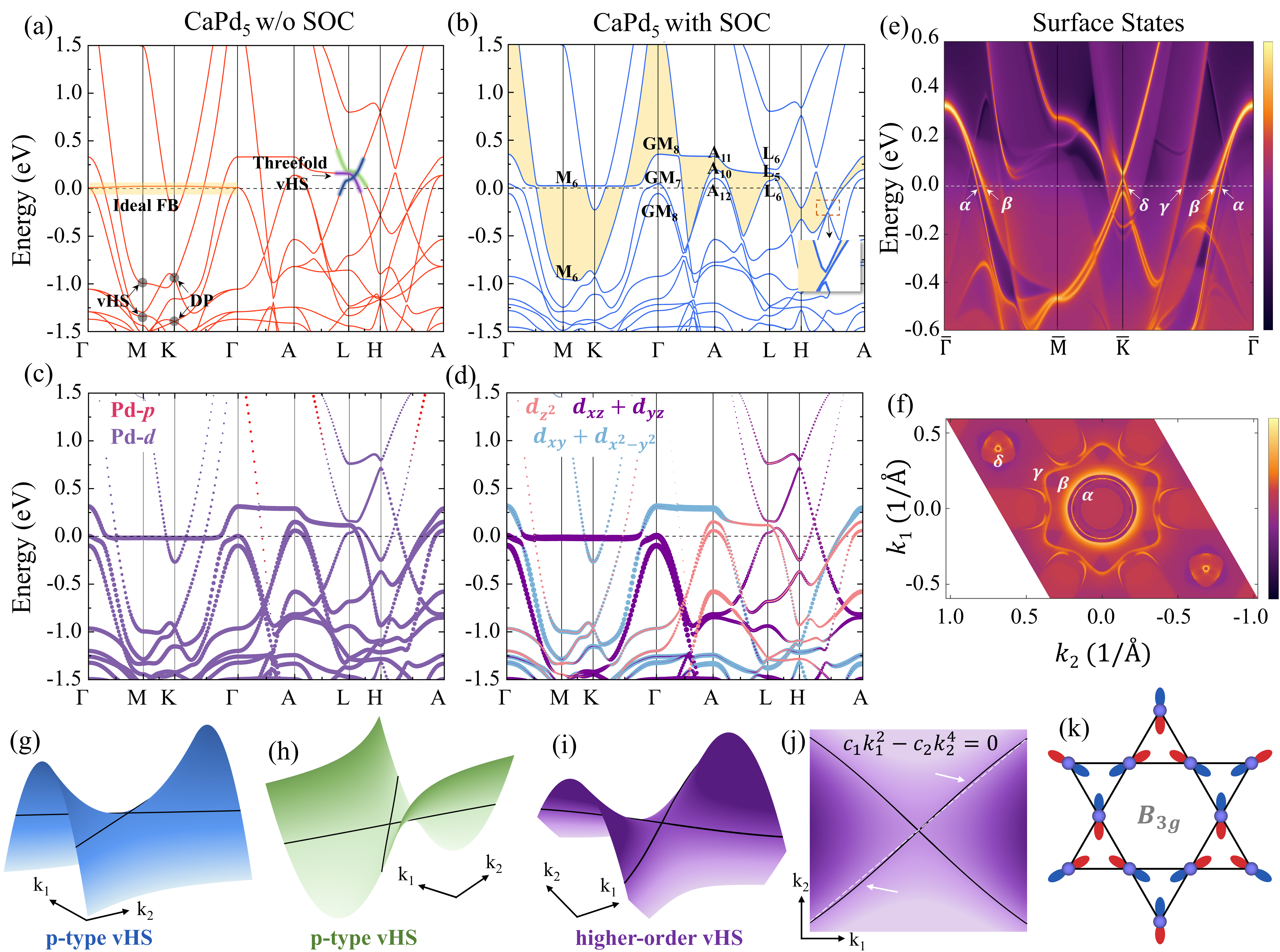}
  \caption{\label{fig2}Electronic structure of CaPd$_5$. (a) Band structure without SOC, where the ideal FB, vHS, and DP are labeled. Notably, at the L point near $E_{\mathrm{F}}$, the solid green and blue lines denote p-type vHS, while the solid purple line indicates a m-type higher-order vHS. (b) Band structure with SOC, the irreducible representations of bands near $E_{\mathrm{F}}$ are labeled. (c) The atomic orbital projected band structure and (d) the orbitally resolved band structure are respectively presented. (e) Surface states along the $\bar{\Gamma}-\bar{\mathrm{M}}-\bar{\mathrm{K}}-\bar{\Gamma}$ high-symmetry path on the (001) surface of CaPd$_5$. (f) Calculated Fermi surface at a fixed energy of 0.002 eV above $E_{\mathrm{F}}$. 3D band structures associated with the vHS identified in panel (a): (g) blue vHS, (h) green vHS, and (i) purple vHS. Black lines and curves delineate the constant energy contours at their respective saddle point energies of 0.224 eV, 0.316 eV, and 0.290 eV for the blue, green, and purple vHS, respectively. (j) Top view of the higher-order vHS 3D band structure, with white dashed lines as guides to confirm its higher-order characteristics. (k) Sign structure (blue or red) and spatial orientation of the $B_{3g}$ orbital at kagome lattice sites.}
\end{figure*}

Topological flat bands are a key characteristic of the kagome lattice, and those at the Fermi level hold particularly rich physical significance\cite{liu2024superconductivity,geng2024correlated}. In the band structure of CaPd$_{5}$ without spin-orbit coupling (SOC), as depicted in Fig. \ref{fig2}(a), a striking feature is the appearance of a flat band with a bandwidth of less than 10 meV, situated precisely at the $E_{\mathrm{F}}$ in the $k_z=0$ plane. Additionally, the $X$Pd$_{5}$ system exhibit threefold vHS at the high-symmetry L point, located approximately 0.3 eV above the Fermi level. The green and blue curves represent p-type van Hove singularities with pure sublattice character, originating from the even parity $d_{xz}$/$d_{yz}$ kagome bands. Their corresponding three-dimensional (3D) band structures are displayed in Figs. \ref{fig2}(g) and \ref{fig2}(h). The purple curve represents a higher-order vHS arising from the odd parity $d_{z^2}$ kagome band with mixed sublattice character, characterized by flat dispersion along L-A directioin.

When SOC is introduced into the system, the flatness of the flat band near the $\Gamma $ point is disrupted, and a band gap opens between the flat band and the adjacent quadratic bands, as shown in Fig. \ref{fig2}. This behavior is a characteristic feature of topological flat bands\cite{bergman2008band,rhim2021singular,yin2022topological}. Similar feature are also observed in the Sr and Ba systems, where the flat bands lie within 0.25 eV above the Fermi level, as illustrated in Fig. S1 of the Supplemental Material. As illustrated in Fig. \ref{fig2}(c), the bands near the Fermi level in CaPd$_{5}$ are primarily derived from Pd-$d$ orbitals. In contrast, the Ca atoms, which do not occupy the kagome sites, exhibit negligible influence on the electronic states near the Fermi level. Fig. \ref{fig2}(d) displays the orbitally resolved band structure of CaPd$_{5}$. Because of the $C_{6z}$ rotational symmetry, the in-plane $d_{xy}$ and $d_{x^2-y^2}$ orbitals are degenerate, as are the out-of-plane $d_{xz}$ and $d_{yz}$ orbitals. Notably, the $d_{xz}$ and $d_{yz}$ orbitals predominantly contribute in the $k_z=0$ plane, a feature linked to their out-of-plane distribution, which impedes in-plane hopping between kagome sites compared to other orbitals. Furthermore, based on the irreducible representations of the site symmetry group $D_{2h}$ at the $3g$ wyckoff position, the dominant $d_{xz}$ and $d_{yz}$ orbitals responsible for forming the flat bands belong to the $B_{3g}$ irreducible representation, with their real-space projections depicted in Fig. \ref{fig2}(k). The orbitally resolved band structures of the Sr and Ba systems exhibit the same features and are shown in Fig. S2.

\begin{table}
  \caption{\label{tab2}%
  Product of parity and $\mathbb{Z} _2 $ indices of the band \#60 in $X$Pd$_{5}$ ($X$=Ca, Sr, Ba). 
  }
  \begin{ruledtabular}
  \begin{tabular}{cccccc}
    \multirow{2}{*}{System}& \multicolumn{4}{c}{Product of parity} & \multirow{2}{*}{$\mathbb{Z} _2 $} \\ 
  & $\Gamma $        & M       & A       & L       &                     \\ \hline
  CaPd$_5$ & 1       & 1       & -1      & 1       & 1                   \\
  SrPd$_5$ & 1       & -1      & -1      & -1      & 1                   \\
  BaPd$_5$ & 1       & -1      & -1      & -1      & 1                  
  \end{tabular}
  \end{ruledtabular}
\end{table}

Since the paramagnetic kagome lattice $X$Pd$_{5}$ possess both time-reversal symmetry and spatial inversion symmetry, and a continuous direct band gap exists between band \#60 and band \#62 at every $k$-point, the $\mathbb{Z} _2$ topological invariant for this pair of bands at the Fermi level can be determined by analyzing the parity of the wavefunctions at the time-reversal invariant momentum (TRIM) points\cite{fu2007topological}. Given that there are three equivalent TRIM points at M and L, one at $\Gamma $ and one at A in the Brillouin zone of $X$Pd$_{5}$ with the $P6/mmm$ space group, we consider only four TRIM points ($\Gamma $, A, M, and L) to determine the $\mathbb{Z} _2$ invariant, as listed in Table \ref{tab2}. The product of the parity eigenvalues of the occupied bands can be calculated using the following equation:
\begin{eqnarray}
  \delta_{i} & = & \Pi_{n} \sqrt{\left\langle\psi_{i, n}\right| \Theta\left|\psi_{i, n}\right\rangle},
\end{eqnarray}
where $\Theta $ is the inversion operator and $\psi_{i, n}$ is the $n$th occupied Bloch function. Then the strong topological invariant $\nu _0$ is given by $(-1) ^{\nu _0} =  \Pi_{i = 1}^{4} \delta_{i}$. This analysis reveals the nontrivial topological properties of the $X$Pd$_{5}$ system, with the Ca, Sr, and Ba systems each hosting topologically flat bands at the Fermi level, characterized by $\mathbb{Z} _2=1$. To further characterize the topological properties of $X$Pd$_{5}$, we labeled the irreducible representations of the bands near the Fermi level, as illustrated in Fig. \ref{fig2}(b) and Fig. S1.

Fig. \ref{fig2}(e) displays the surface states of CaPd$_{5}$ around the flat band. As a topological metal with a complex band structure near the Fermi level, the system exhibits intricate surface states below the Fermi energy. In particular, the prominent surface states around the $\Gamma$ point are topologically nontrivial, stemming from the band inversion around the A point (see Supplemental Material Fig. S3). Moreover, characteristic Dirac-cone surface state emerges at the $M$ point at around -0.5 eV, with SrPd$_{5}$ and BaPd$_{5}$ exhibiting the same feature, as shown in Fig. S4. Furthermore, we calculated the Fermi surface at a fixed energy of 0.002 eV above the Fermi level, as shown in Fig. \ref{fig2}(f). In addition, the corresponding contributing bands within the surface states were identified and labeled as $\alpha $, $\beta $, $\gamma $ and $\delta $, respectively.

\subsection{Topological flat-band of the bosonic mode}

\begin{figure*}
  \includegraphics[width=0.9\textwidth]{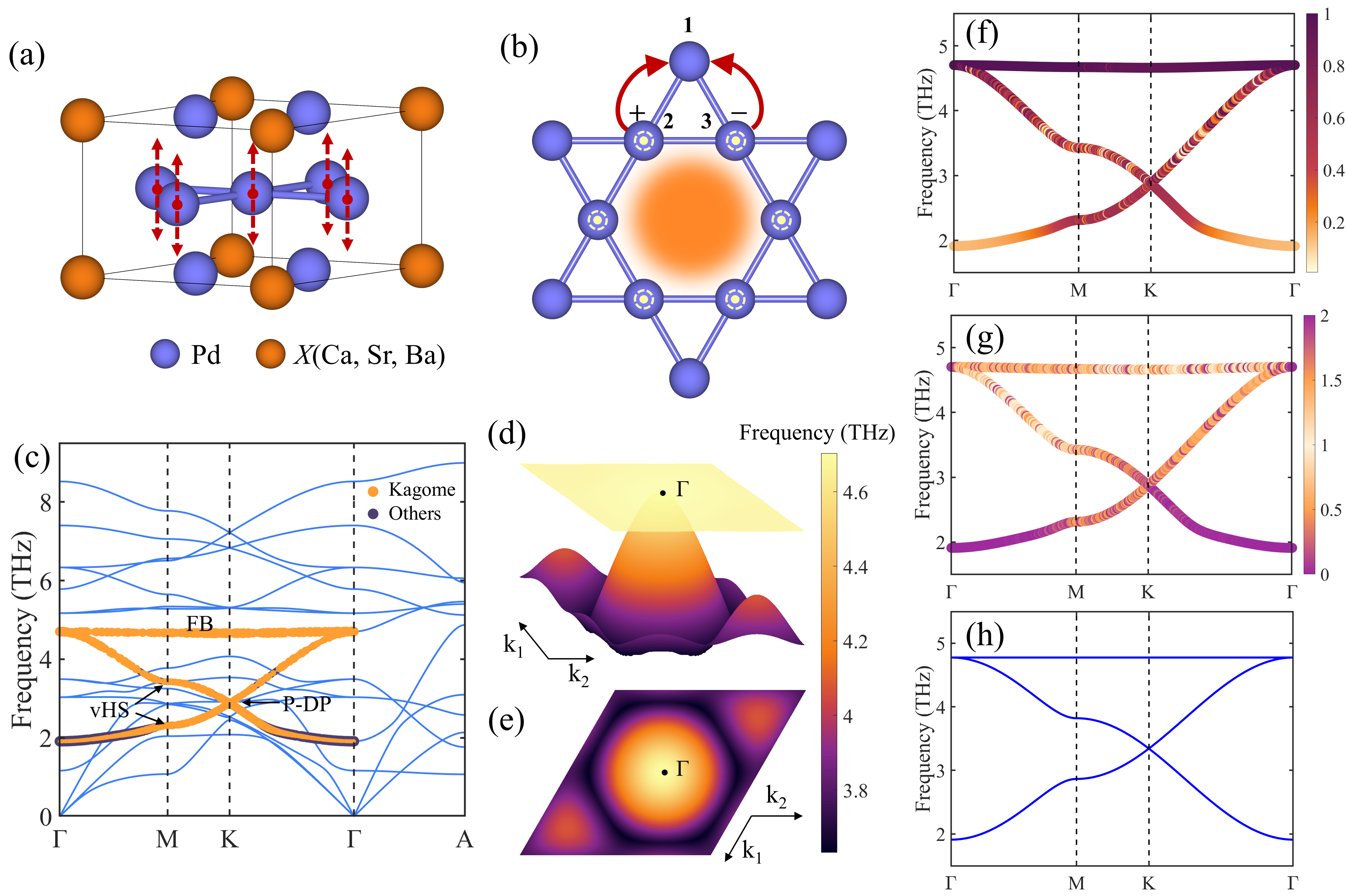}  
  \caption{\label{fig3}Kagome flat-band origin of the bosonic mode in CaPd$_5$. (a) Calculated atomic displacements associated with the phonon flat-band in CaPd$_5$,  with red arrows indicating the directions of atomic vibrations. (b) Top view of the kagome lattice with three kagome sites labeled as 1, 2, and 3, where $+$ and $-$ represent the vibrational phases of the flat band at neighboring sublattices, respectively. (c) Atomic contribution projection of the kagome phonon flat-band, where orange indicates contributions from Pd atoms on the kagome lattice, and dark blue represents contributions from other atoms. (d) Side view and (e) bottom view of the 3D phonon flat-band. (f) The proportion of $z$-direction vibrations at kagome sites relative to total atomic vibrations. (g) Phase difference of vibrations between sites 2 and 3. (h) Phonon spring-mass model of CaPd$_5$.}.
\end{figure*}

The behavior of bosons in the system is often associated with novel physical phenomena such as superconductivity and charge density waves \cite{yin2022topological}. Therefore, we conducted a more in-depth investigation into the phonon vibrational modes of $X$Pd$_{5}$. Given the similar phonon behavior across the system, we focus on CaPd$_{5}$ as a representative example, with its phonon spectrum shown in Fig. \ref{fig3}(c). Interestingly, the phonon spectrum of this system exhibits features analogous to those of the classic kagome electronic band structure, including a flat band, van Hove singularities, and a Dirac point. Due to the absence of spin degrees of freedom in bosons, the crossing point in the phonon spectrum is two-fold degenerate, in contrast to the four-fold degenerate DP in electronic systems. We refer to these as pseudo-Dirac points (P-DP). Additionally, the calculated 3D phonon flat band exhibits a typical quadratic band touching with its adjacent bands at the $\Gamma $ point, as illustrated in Figs. \ref{fig3}(d-e). To further confirm the origin of the kagome-characteristic phonon spectrum (KCPS), we performed atomic projections on this portion of the spectrum, as shown in Fig. \ref{fig3}(c). The result indicates that the KCPS is primarily contributed by Pd atoms occupying the kagome sites.

To elucidate how the $X$Pd$_{5}$ systems give rise to the KCPS, we analyzed the vibrational modes of the atoms occupying the kagome sites. Generally speaking, the atoms in the KCPS vibrate along the out-of-plane direction, as indicated by the red arrows in Fig. \ref{fig3}(a). A more detailed description of this dynamic atomic vibration is provided in Fig. \ref{fig3}(f), which illustrates the proportion of vibrations along the $z$-direction at the kagome sites relative to the total vibrations, denoted as $A_{\mathrm{kg},z}/A_{\mathrm{tot}}$. Here, $A_{\mathrm{kg},z}$ represents the total amplitude of vibrations along the $z$-direction for atoms at the kagome sites, while $A_{\mathrm{tot}}$ represents the total amplitude of vibrations along the $x$, $y$, and $z$-directions for all atoms. Interestingly, on the phonon flat band, the proportion of vibration $A_{\mathrm{kg},z}/A_{\mathrm{tot}}$ is nearly 1, indicating that the flat band originates from atoms vibrating along the $z$-direction, distinct from the more common contributions arising from collective lattice displacements within a hexagonal ring of the kagome lattice\cite{yin2020fermion,yin2022topological}. The construction of in-plane phonon flat bands from out-of-plane vibrations is a novel phenomenon that has recently been predicted theoretically\cite{xia2024negative}.

The phase difference in the vibrations of kagome atoms may provide further insights into the mechanism by which out-of-plane vibrations construct in-plane phonon flat bands. Fig. \ref{fig3}(g) shows the vibrational phase difference between kagome sites 2 and 3 in Fig. \ref{fig3}(b), denoted as $|\boldsymbol{\phi}_{2,z}-\boldsymbol{\phi}_{3,z}|$, where $\boldsymbol{\phi}_{2,z}$ and $\boldsymbol{\phi}_{3,z}$ represent the vibration phases along the $z$-direction for sites 2 and 3, respectively. The zero vibrational phase difference indicates that the vibrations at the two sites are in phase, while a vibrational phase difference of 1 signifies that the vibrations are exactly out of phase. Here, we can draw an analogy to the mechanism underlying the formation of electronic flat bands, where electronic eigenstates with opposite phases exist on the hexagon. Any hoppings outside the hexagon are canceled by destructive quantum interference, leading to the perfect localization of electrons\cite{bergman2008band,kang2020topological,regnault2022catalogue}. In $X$Pd$_{5}$, the out-of-plane phonon vibrations can be analogously compared to the $p_z$ electronic orbitals, while the phonon flat bands arise from the destructive interference of phonons with opposite vibrational phases on the hexagon. These dispersionless phonon flat bands arise from interference effects and are topological in nature. Their localized character is demonstrated by strictly localized eigenfunctions in real space, known as compact localized states\cite{bergman2008band,rhim2019classification,xia2024negative}, as illustrated in the orange region of Fig. \ref{fig3}(b).

To further deepen our understanding of the phonon behavior in this system, we constructed a phonon spring-mass model and solve the eigenvalue of the dynamical matrix $D(\mathbf{q})$ to obtain the dynamical properties of the atoms within the harmonic approximation\cite{togo2015first}:
\begin{eqnarray}
  \mathrm{D}(\mathbf{q}) \mathbf{e}_{\mathbf{q} j} = \omega_{\mathbf{q} j}^{2} \mathbf{e}_{\mathbf{q} j},
\end{eqnarray}
where $\mathbf{q}$ is the wave vector and $j$ is the band index, $\omega _{\mathbf{q}j}$ and $e _{\mathbf{q}j}$ give the phonon frequency and polarization vector of the phonon mode labeled by a set ${\mathbf{q},j}$, respectively. And the dynamical matrix $D(\mathbf{q})$ is:
\begin{eqnarray}
  D_{\alpha \beta}^{m n} & = & \sum_{R_L} \frac{1}{\sqrt{M_{m} M_{n}}}\left(-\Phi_{\alpha \beta}^{m n}(R_L)\right) e^{i \boldsymbol{q} \cdot\left(\boldsymbol{\tau }_{n}-\boldsymbol{\tau }_{m}-R_L\right)}
\end{eqnarray}
Here, $m$ and $n$ are the atom indices, $R_L$ is the lattice vector, $\boldsymbol{\tau } $ denotes the relative atomic position, and $\Phi_{\alpha \beta}^{m n}(R)$ represents the second-order force constants. For a system where atoms vibrate along the $z$-direction, the force constant matrix $\Phi_{\alpha \beta}^{m n}(R)$ can be simplified to the spring constant $k^f$\cite{xia2024negative}. Within this model, the interatomic spring constant is set to $k^f _{mn}=0.21(m\ne n)$, while the on-site spring constant is $k^{f} _{mm}=0.42$. $M$ is the mass of Pd atom, which is set to 106.42. By considering the 4 nearest-neighbor sites, we obtain the phonon spring-mass model as shown in Fig. \ref{fig3}(h).

\subsection{Electron-phonon coupling and superconducting state}

\begin{figure}
  \includegraphics[width=\columnwidth]{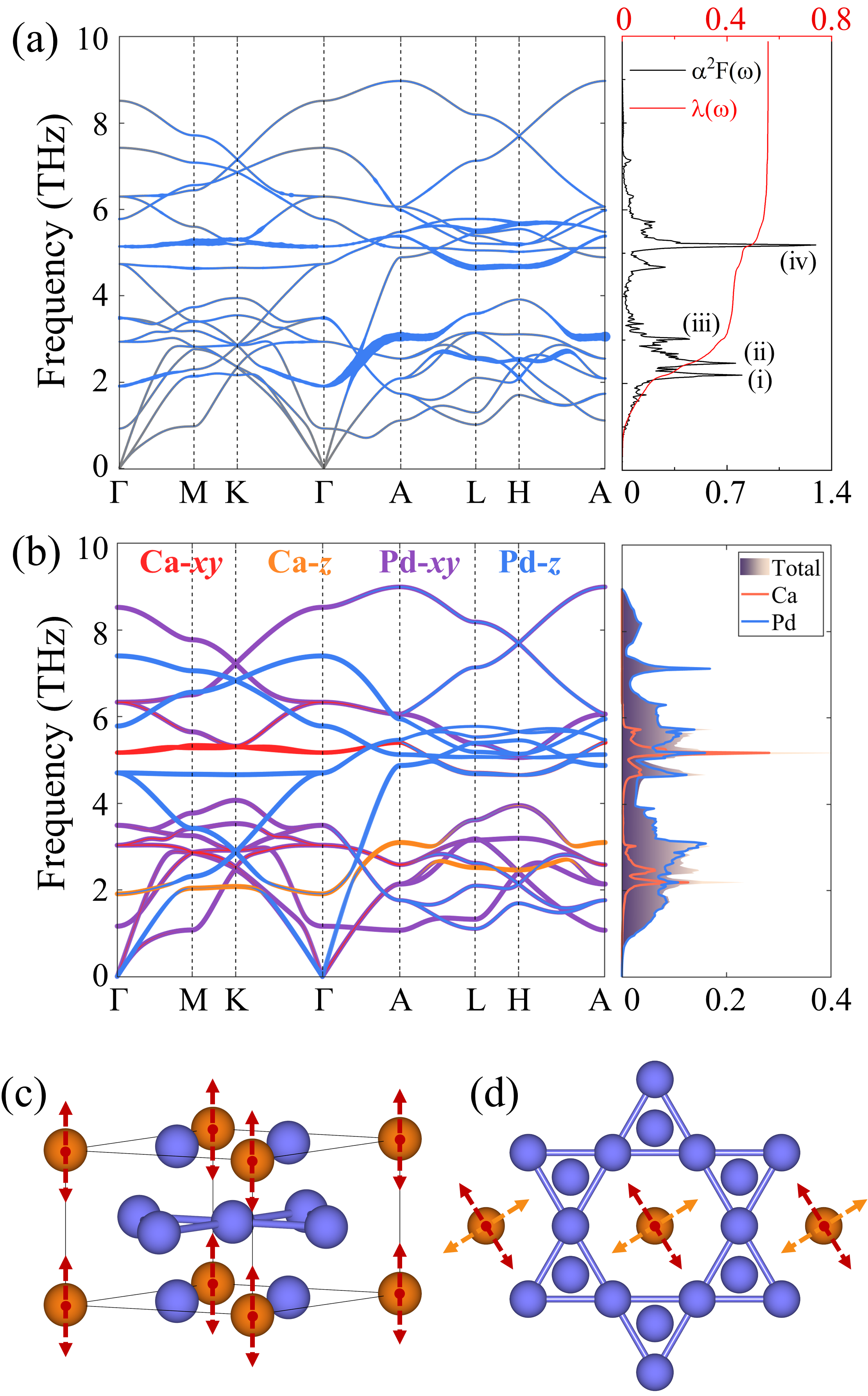}
  \caption{\label{fig4} (a) Phonon dispersions weighted by the magnitude of the phonon linewidth for CaPd$_5$. The right panel is the Eliashberg spectral function $\alpha ^2F(\omega )$ (black line), and the integrated strength of EPC $\lambda (\omega )$ (red line). (b) Phonon dispersion weighted by different atomic vibrational modes of CaPd$_5$. The right panel is the total (colored zone) and vibrational mode-resolved (colored lines) phonon density of states. Figs. (c)–(d) illustrate the atomic vibrations corresponding to the principal peaks in the $\alpha ^2F(\omega )$ function shown in Fig. (a). Specifically, Fig. (c) depicts the primary atomic vibrations for peaks \romannumeral1, \romannumeral2, and \romannumeral3, while Fig. (d) shows for peak \romannumeral4.}
\end{figure}

To further investigate the effects of electron-phonon interactions in the system, we calculated the electron-phonon coupling in $X$Pd$_{5}$ using QE. According to the isotropic Migdal-Eliashberg theory\cite{allen1983theory,margine2013anisotropic}, the strength of the EPC, denoted by $\lambda_{\mathbf{q} \nu}$, can be evaluated as follows:
\begin{eqnarray}
  \lambda_{\mathbf{q} \nu} & = & \frac{\gamma_{\mathbf{q} \nu}}{\pi h N\left(E_{\mathrm{F}}\right) \omega_{\mathbf{q} \nu}^{2}} .
  \label{eqnumda}
\end{eqnarray}
Here, $N\left(E_{F}\right)$ is the density of states at the Fermi level, $\omega_{\mathbf{q} \nu}$ represents the phonon frequency of the $\nu$th phonon mode with wave vector $\mathbf{q}$, and $\gamma_{\mathbf{q} \nu}$ denotes the phonon linewidth, defined as :
\begin{align}
	\gamma_{\mathbf{q} \nu} &= \frac{2 \pi \omega_{\mathbf{q} \nu}}{\Omega_{\mathrm{BZ}}}
	\sum_{\mathbf{k}, \mathbf{n}, \mathbf{m}}\left|g_{\mathbf{k} \mathbf{n}, \mathbf{k}+\mathbf{q} \mathbf{m}}^{\nu}\right|^{2} \nonumber \\
	&\quad \times \delta\left(\varepsilon_{\mathbf{k} \mathbf{n}}-E_{F}\right) \delta\left(\varepsilon_{\mathbf{k}+\mathbf{q} \mathbf{m}}-E_{F}\right),
\end{align}
where $\Omega_{\mathrm{BZ}}$ is the volume of the Brllouin zone (BZ), $\mathbf{n}$ and $\mathbf{m}$ are the Kohn-Sham orbitals, $\varepsilon_{\mathbf{k} \mathbf{n}}$ and $\varepsilon_{\mathbf{k}+\mathbf{q} \mathbf{m}}$ are the Kohn-Sham energies, $E_{F}$ is the Fermi energy, and $g_{\mathbf{k} \mathbf{n}, \mathbf{k}+\mathbf{q} \mathbf{m}}^{\nu}$ corresponds to the EPC matrix element. Furthermore, the total electron-phonon coupling $\lambda(\omega)$ can be calculated by
\begin{eqnarray}
\lambda(\omega)=2 \int_{0}^{\omega} \frac{\alpha^{2} F(\omega)}{\omega} d \omega,
\end{eqnarray}
where $\alpha^{2} F(\omega)$ represents the Eliashberg electron-phonon spectral function, which is defined as:
\begin{eqnarray}
  \alpha^{2} F(\omega)=\frac{1}{2\pi N\left(E_{F}\right)} \sum_{\mathbf{q} \nu} \delta\left(\omega-\omega_{\mathbf{q} \nu}\right) \frac{\gamma_{\mathbf{q} \nu}}{\omega_{\mathbf{q} v}}.
  \label{a2f}
\end{eqnarray}
Fig. \ref{fig4}(a) presents the phonon linewidths of the CaPd$_5$ system alongside the calculated $\alpha^{2} F(\omega)$ and $\lambda(\omega)$. The Eliashberg spectral function $\alpha^{2} F(\omega)$ exhibits prominent peaks centered at around $2\thicksim 3$ THz (labeled as \romannumeral1, \romannumeral2, \romannumeral3) and $\thicksim 5.3$ THz (labeled as \romannumeral4), which primarily govern the accumulation of the EPC strength. Our calculations yield EPC constants of $\lambda=0.557$ for CaPd$_5$, with SrPd$_5$ and BaPd$_5$ exhibiting $\lambda=0.494$ and $0.559$, respectively.

According to the Eqs. \ref{eqnumda} and \ref{a2f}, the softening phonon modes can enhance the EPC strength and the values of $\alpha^{2} F(\omega)$. As shown in Fig. \ref{fig4}(a), the low-frequency peaks in the $\alpha^{2} F(\omega)$ originates from softening phonon modes along the $\Gamma $-A path and near the A point, while the middle-frequency peak arises from softening modes confined to the $k_z=0$ plane. Comparative analysis with Fig. \ref{fig4}(b) reveals the vibrational origins: the low-frequency softening modes predominantly derive from Ca-$z$ vibrations, whereas the middle-frequency modes emerge from Ca-$xy$ vibrations. The schematic illustrations of these atomic vibrations that enhance the electron-phonon coupling are further presented in Figs. \ref{fig4}(c)-(d). In the low-frequency region (peaks \romannumeral1, \romannumeral2, \romannumeral3), the Ca atoms vibrate along the out-of-plane direction, while in the middle-frequency region (peak \romannumeral4), they vibrate along two mutually perpendicular directions within the $xy$-plane, as indicated by the red and orange arrows, respectively. Since the electronic states near the Fermi level are predominantly contributed by Pd atoms, the interaction with the vibrations of Ca atoms at the hexagonal centers can establish an effective Holstein-type electron-phonon coupling for Pd atoms in kagome lattice\cite{wu2024crossover}. This effective electron-phonon interaction would significiantly contribute to the superconducting pairing.

\begin{figure}
  \includegraphics[width=\columnwidth]{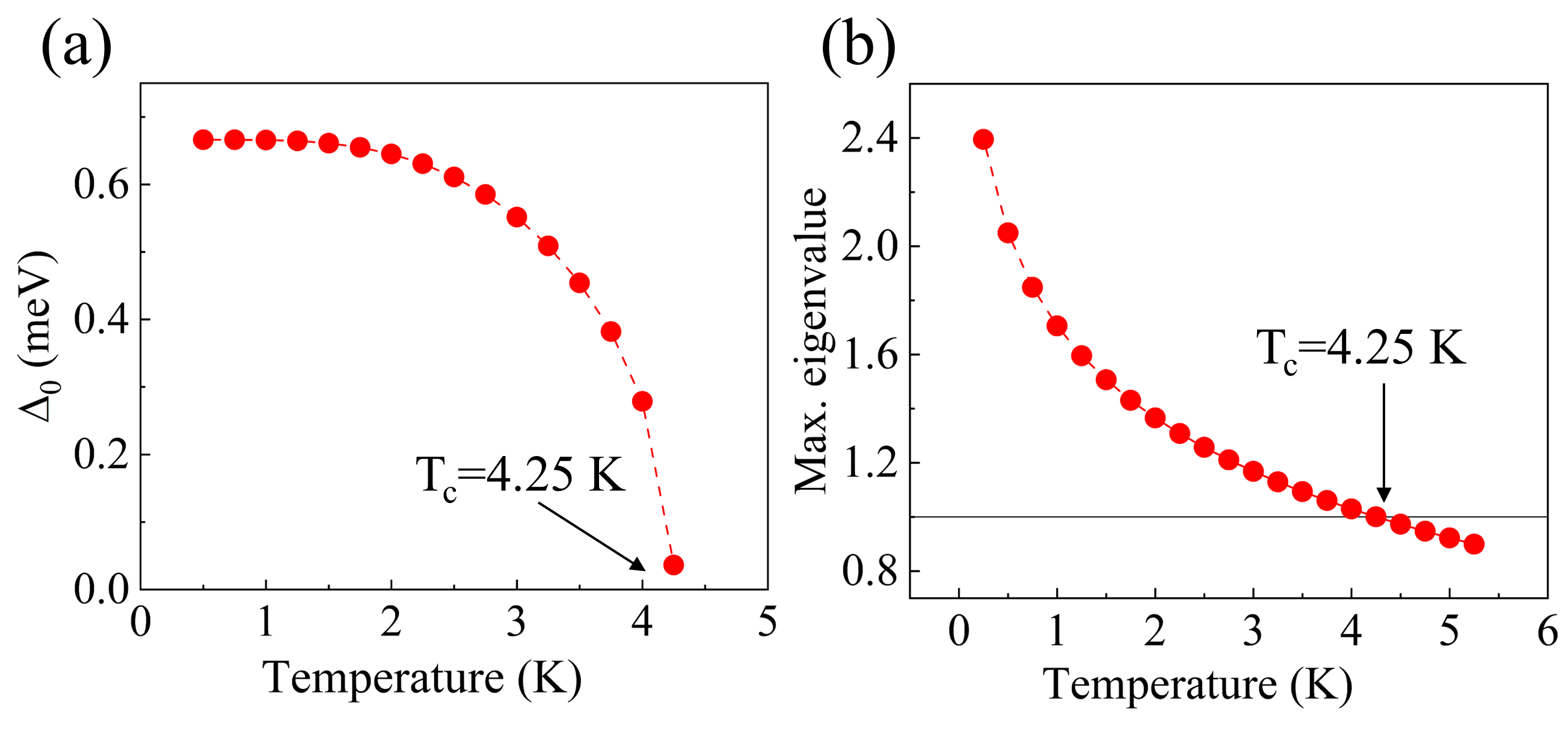}
  \caption{\label{fig5} Calculated (a) superconducting gap and (b) maximum eigenvalue of CaPd$_5$ as a function of temperature.}
\end{figure}

Based on the EPC behavior studied, we can further explore the superconducting state of the $X$Pd$_{5}$ system. The superconducting gap $\Delta$ of the system can be obtained by solving the isotropic Migdal-Eliashberg equations:
\begin{eqnarray}
  Z\left(i \omega_{j}\right) \Delta\left(i \omega_{j}\right)=\pi T \sum_{j^{\prime}} \frac{\Delta\left(i \omega_{j^{\prime}}\right)}{\sqrt{\omega_{j^{\prime}}^{2}+\Delta^{2}\left(i \omega_{j^{\prime}}\right)}} \nonumber \\
  \times \left[\lambda\left(\omega_{j}-\omega_{j^{\prime}}\right)-\mu_{\mathrm{c}}^{*}\right].
  \label{ME-eq}
\end{eqnarray}
Here, $\mu_{\mathrm{c}}^{*}$ is the Coulomb parameter, which is set to 0.1 in the $X$Pd$_{5}$ calculations, and $i \omega{j} = i(2j+1) \pi T$ ($j$ is an integer) represents the fermion Matsubara frequencies. As the temperature approaches $T_c$, $\Delta\left(i \omega_{j}\right)$ tends to zero, as shown in Fig. \ref{fig5}(a). Meanwhile, the set of Eq. (\ref{ME-eq}) can be transformed into a linear equation for $\Delta\left(i \omega_{j}\right)$:
\begin{align}
	& \Delta\left(i \omega_{j}\right) = \sum_{j^{\prime}} K_{j j^{\prime}} \Delta\left(i \omega_{j^{\prime}}\right), \\
	K_{j j^{\prime}} &= \frac{1}{\left|2 j^{\prime}+1\right|} \bigg[ \lambda\left(\omega_{j} - \omega_{j^{\prime}}\right) - \mu_{\mathrm{c}}^{*} \nonumber \\
	&\quad - \delta_{j j^{\prime}} \sum_{j^{\prime \prime}} \lambda\left(\omega_{j}-\omega_{j \prime \prime}\right) s_{j} s_{j^{\prime \prime}} \bigg],
\end{align}
where $s_{j}=sign\left(\omega_{j}\right)$. The critical temperature $T_c$ is defined as the valueat which the maximum eigenvalue of $K_{j j^{\prime}}$ is 1\cite{allen1983theory,margine2013anisotropic}, as shown in Fig. \ref{fig5}(b). bur calculations reveal superconducting transition temperatures of 4.25 K for CaPd$_5$, with SrPd$_5$ and BaPd$_5$ exhibiting $T_c=2.75$ K and $3.35$ K, respectively, as demonstrated in the supplementary Figs. S5 and S6.

\section{CONCLUSIONS}
In summary, we propose that the kagome metals $X$Pd$_{5}$ ($X$=Ca, Sr, Ba) serve as a rich arena for investigating the interplay between localized fermions and bosons and the superconducting states. Our results demonstrate that $X$Pd$_{5}$ hosts flat bands near the Fermi level with a nontrivial $\mathbb{Z} _{2}=1$ topological invariant. We identified multifold vHS near the Fermi level, including two p-type vHS and a higher-order vHS. Remarkably, the in-plane phononic flat bands arising from out-of-plane atomic vibrations exhibit a novel origin: destructive interference between adjacent kagome lattice sites with antiphase vibrational modes. This unambiguously demonstrates that geometric frustration induces phonon flat bands in the system. First-principles electron-phonon coupling calculations confirm superconducting ground states in the $X$Pd$_{5}$ series, with predicted transition temperatures $T_c=4.25$ K, $2.75$ K, and $3.35$ K for CaPd$_{5}$, SrPd$_{5}$ and BaPd$_{5}$, respectively. This work provides three stable superconducting materials to explore fermion-boson many-body physics, potentially advancing discoveries in charge density waves and unconventional superconductivity. More importantly, we propose a generalized framework for identifying geometric frustration-induced topological phononic flat bands, providing systematic insights into their origin.

\begin{acknowledgments}
We acknowledge the financial support by National Natural Science Foundation of China (Nos. 11874113, 62474041, 12274078), Natural
Science Foundation of Fujian Province of China (Nos. 2020J02018, 2023J01519). X. W. is supported by the National Key R\&D Program of China (Grant No. 2023YFA1407300) and the National Natural Science Foundation of China (Grant No. 12447103).
\end{acknowledgments}





\bibliography{apssamp}

\end{document}